\documentclass{article}

\usepackage{microtype}
\usepackage{graphicx}
\usepackage{booktabs}
\usepackage{hyperref}

\usepackage[preprint]{icml2026}
\usepackage{amsmath}
\usepackage{amssymb}
\usepackage{multirow}
\usepackage{stfloats}
\usepackage{xcolor}

\newif\ifaffilsunderauthors
\affilsunderauthorstrue
\makeatletter
\newcommand{\affiliationsblock}{%
  \ifaffilsunderauthors
    {\setlength\topsep{0pt}\setlength\parskip{0pt}%
    \vskip 0.1in
    \begin{center}\small
      \@tempcnta=\z@
      \loop\ifnum\@tempcnta<\value{@affiliationcounter}%
        \advance\@tempcnta\@ne
        \ifnum\@tempcnta>\@ne\hspace{2.5em}\fi
        \textsuperscript{\the\@tempcnta}\,\csname @affilname\the\@tempcnta\endcsname
      \repeat
    \end{center}}%
  \fi}
\newcommand{\printaffiliationsfootnote}{%
  \ifaffilsunderauthors
    \global\icml@noticeprintedtrue
    {\let\thefootnote\relax\footnotetext{\hspace*{-\footnotesep}{\raggedright\hyphenpenalty=10000\exhyphenpenalty=10000 Correspondence to: \icmlcorrespondingauthor@text.\par}\Notice@String}}%
  \else
    \printAffiliationsAndNotice{}%
  \fi}
\makeatother

\icmltitlerunning{SATFRONT: Detecting Atmospheric Fronts with Satellite Data using Machine Learning}

\begin{document}

\twocolumn[
  \icmltitle{SATFRONT: Detecting Atmospheric Fronts with\\Satellite Data using Machine Learning}

  \begin{icmlauthorlist}\ifaffilsunderauthors\setlength\baselineskip{15pt}\fi
    \icmlauthor{Ioannis Dravilas}{uva}
    \icmlauthor{Miruna Stoicescu}{eum}
    \icmlauthor{Roope Tervo}{eum}
    \icmlauthor{Carlos Horn}{eum}\\
    \icmlauthor{Oriol Hinojo Comellas}{eum}
    \icmlauthor{Sina Montazeri}{eum}
    \icmlauthor{Andrea Meraner}{eum}
    \icmlauthor{Ana Lucic}{uva}
  \end{icmlauthorlist}

  \icmlaffiliation{uva}{University of Amsterdam, Amsterdam, The Netherlands}
  \icmlaffiliation{eum}{EUMETSAT, Darmstadt, Germany}
  \affiliationsblock

  \icmlcorrespondingauthor{Ioannis Dravilas}{giannisdravilas@gmail.com}

  \icmlkeywords{atmospheric fronts, satellite remote sensing, semantic segmentation, deep learning}

  \vskip 0.3in
]

\printaffiliationsfootnote

\begin{abstract}
Atmospheric fronts are associated with much of the high-impact weather in the mid-latitudes, yet their identification and classification remain largely manual and subjective. Existing deep learning approaches rely primarily on gridded reanalysis or forecast data, limiting their applicability to satellite observations available for time-critical applications. We introduce SATFRONT, a machine learning framework for detecting cold, warm, and occluded fronts over Europe from geostationary satellite observations, with reanalysis data as an optional additional input. The model is trained on frontal analyses from five European national meteorological services and evaluated against UK Met Office analyses for 2024. Using satellite observations alone, SATFRONT achieves a Critical Success Index (CSI) of 0.428 at a 150\,km tolerance, within the range of agreement between analyses from different meteorological services (0.277--0.501) and higher than that of an operational rule-based front detector that uses satellite data (0.133). Adding ERA5 reanalysis increases the CSI to 0.591. ERA5 substantially improves the placement of individual fronts, although its effect on their annual spatial distribution varies by front type. These results demonstrate that cold, warm, and occluded fronts can be detected directly from geostationary satellite observations, providing a basis for automated frontal analyses at the high temporal frequency required for time-critical applications.
\end{abstract}

\section{Introduction}
\label{sec:introduction}

\begin{figure}[t]
\centering
\includegraphics[width=0.85\columnwidth]{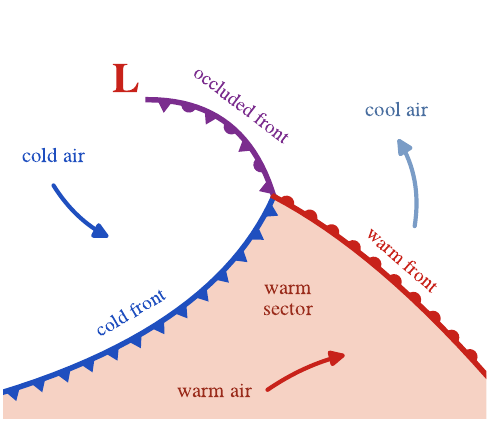}
\caption{Schematic of a low-pressure system (L) with a cold front (blue, triangles), a warm front (red, semicircles), and an occluded front (purple, alternating symbols). The symbols point in the direction in which each front moves. The shaded area between the cold and the warm front is the warm sector, and arrows indicate the motion of the air masses.}
\label{fig:schematic}
\end{figure}

Atmospheric fronts are boundaries between air masses with different thermodynamic properties and are associated with a large fraction of precipitation and severe weather in the mid-latitudes \citep{Catto2013-em, Berry2011-jq}. Cold, warm, and occluded fronts are therefore routinely identified in operational weather analysis. A cold front is the leading edge of advancing cold air, a warm front is the leading edge of advancing warm air, and an occluded front forms where a cold front overtakes a warm front \citep{Hewson1998-hg} (Figure~\ref{fig:schematic}). In many national meteorological services, however, frontal analyses are still produced predominantly by forecasters drawing fronts manually on surface analysis charts. This process is time-consuming and inherently subjective, as different services analysing the same weather situation may disagree on the position, continuity, and classification of fronts \citep{Justin2025-ti}. Automated methods could provide forecasters with a reproducible first-guess analysis that can subsequently be edited \citep{Justin2025-ti}, while also enabling consistent large-scale and long-term studies of frontal activity \citep{Berry2011-jq, Rudeva2015-dx}. However, machine-readable archives of frontal analyses, which are needed to train and evaluate such methods \citep{Niebler2022-ih}, remain scarce outside North America.

Most automated front-detection methods rely on fully gridded atmospheric fields. Traditional approaches identify fronts from horizontal gradients of temperature, humidity, or related quantities using predefined physical criteria \citep{Hewson1998-hg, Schemm2015-xu, Berry2011-jq}. More recently, studies using machine learning and deep learning methods have shown improved performance on automated front detection \citep{Biard2019-to, Lagerquist2019-bg, Bochenek2021-nl, Justin2025-ti}, but so far depend on atmospheric state variables from numerical weather prediction (NWP) or reanalysis products. NWP models, such as the Integrated Forecasting System \citep{ECMWF2023-ifs}, simulate the atmosphere by solving the equations of fluid dynamics on a grid, while reanalysis products such as ERA5 \citep{Hersbach2020-or} combine historical observations with an NWP model to reconstruct past states of the atmosphere. Operational NWP analyses, which estimate the current state of the atmosphere, are typically available only a few hours after the time they describe, and the forecasts contain no observational information after their initialisation time. Reanalysis data are particularly useful for retrospective studies and model development, but they are produced with a delay of several days and are therefore not suitable as an input to a system intended for time-critical operation.

In contrast, geostationary satellites observe the atmosphere multiple times per hour, hence their data are available at a frequency suitable for time-critical operational forecasting. Infrared satellite channels capture large-scale cloud structures, cloud-top features, moisture patterns, and other features associated with frontal systems, making them a promising source of information for automated frontal analysis. 
However, satellite observations do not directly provide the temperature, humidity, and wind variables at scales and vertical levels conventionally used to identify fronts.

Despite this operational advantage, there are not many methods for satellite-based front detection. The operational Automatic Satellite Image Interpretation (ASII) product detects fronts from Meteosat Second Generation observations using manually designed rules \citep{NWCSAF2022-asii}. The only previous deep learning method that incorporates satellite observations, Cloud-DETR \citep{Qin2024-detr}, detects only cold fronts over an Asian domain and additionally requires ERA5 data. To our knowledge, no previous deep learning system has demonstrated detection of cold, warm, and occluded fronts directly from satellite observations alone.

In this work, we introduce \textbf{SATFRONT}, a deep learning framework for detecting cold, warm, and occluded fronts over Europe from geostationary satellite observations. We consider two input configurations: a satellite-only configuration designed to assess whether fronts can be detected without gridded atmospheric state variables from models, and a satellite+ERA5 configuration that additionally includes ERA5 data. The satellite+ERA5 configuration allows us to quantify how much information ERA5 contributes beyond that already present in the satellite observations.

A further fundamental challenge in this task is that frontal analyses from different meteorological services follow different conventions. Rather than selecting a single source of labels, we train on analyses from five European national meteorological services. SATFRONT uses a shared segmentation network with a separate final layer for each service, allowing the model to learn a common representation from all available training data while retaining separate predictions for each service's analysis conventions.

Our main contributions are:
\begin{enumerate}
    \item We introduce, to our knowledge, the first deep learning model that detects cold, warm, and occluded fronts from geostationary satellite observations alone. Evaluated against UK Met Office analyses, our satellite-only model achieves a Critical Success Index (CSI) within the range of agreement observed between different meteorological services and substantially outperforms an operational rule-based front detector that uses satellite data.

    \item We develop a multi-service training strategy that combines frontal analyses from five European meteorological services through a shared network with a separate final layer for each service. This allows our model to benefit from a larger set of labelled analyses without requiring the services to follow identical frontal analysis conventions.

    \item We compare satellite-only and satellite+ERA5 configurations using the same model architecture, isolating the contribution of gridded atmospheric information. Adding ERA5 substantially improves the placement of fronts in individual analyses.

    \item We evaluate the annual spatial distribution of frontal activity. This analysis shows that additional ERA5 information affects short-term front placement and long-term frontal frequency statistics differently between front types.
\end{enumerate}

Together, these experiments test whether frontal analysis can be performed directly from high-frequency geostationary satellite observations and clarify which aspects of the problem still benefit from information contained in reanalysis data.

\section{Related Work}
\label{sec:related_work}

\subsection{Automated front detection}

Automated front detection has traditionally been based on diagnostic quantities derived from gridded atmospheric fields. A widely used family of methods follows the framework of \citet{Hewson1998-hg}, in which fronts are identified from strong horizontal gradients in temperature or related thermodynamic variables. Subsequent studies have adapted these diagnostics to different datasets and applications, including global climatologies of frontal activity \citep{Berry2011-jq, Schemm2015-xu, Rudeva2015-dx}. Because these approaches apply explicit thresholds to physical quantities, it is clear why they detect a front at a given location, but their performance depends on predefined thresholds and on the availability of gridded atmospheric state variables.

More recent work has replaced such manually designed detection criteria with supervised machine learning models. \citet{Lagerquist2019-bg} trained convolutional neural networks on North American reanalysis data to detect cold and warm fronts, using fronts analysed by forecasters at the US Weather Prediction Center as targets. \citet{Bochenek2021-nl} used a random forest method on ERA5 data over Central Europe using fronts digitised from German Weather Service charts, with all front types merged into a single class. \citet{Biard2019-to} used a convolutional neural network to detect fronts from multiple surface variables over North America. More recently, FrontFinder~AI \citep{Justin2025-ti} applied a U-Net-based architecture \citep{Ronneberger2015-me} to ERA5 data to identify cold, warm, stationary, and occluded fronts over the United States and surrounding regions.

Despite differences in architecture and target definition, these machine learning methods share an important characteristic: they operate on gridded atmospheric variables from reanalysis or similar products, which become available only hours to days after the observations. These methods therefore cannot detect fronts as soon as new observations arrive. Our work addresses a different setting by asking whether the same frontal structures can be inferred directly from satellite observations, which are available with low latency but in which variables that are critical for detecting fronts, such as temperature, humidity, pressure, and wind, are not explicitly available.

\subsection{Automated front detection from satellite observations}

Satellite observations have long been used by forecasters to identify frontal cloud structures, but comparatively few automated methods operate directly on satellite observations. The ASII product \citep{NWCSAF2022-asii}, developed within the Nowcasting Satellite Application Facility (NWC SAF) of the European Organisation for the Exploitation of Meteorological Satellites (EUMETSAT), detects cold, warm, and occluded fronts from Meteosat Second Generation observations using manually designed rules. These rules combine information from infrared channels, channel differences, and characteristics of cloud structure and motion. An additional mode incorporates NWP variables to refine the detected fronts and their classifications.

Deep learning approaches using satellite observations are considerably more limited. Cloud-DETR \citep{Qin2024-detr} applies a transformer-based object detector to satellite observations from the Fengyun satellites \citep{Zhang2019-fengyun} together with two variables derived from ERA5. It detects cold fronts over an Asian domain, but does not consider warm or occluded fronts and does not operate from satellite observations alone. The authors do not explicitly justify their decision to exclude warm and occluded fronts, nor do they discuss whether their method could be extended to these front types, so the applicability of Cloud-DETR to other front types remains unexplored. In addition, the cold fronts used as labels were identified manually from ERA5 fields, such as the temperature advection at 850\,hPa and the sea level pressure, and two of the three input channels of the model are derived from the same ERA5 fields. Because the inputs and the labels are not independent, it is difficult to assess how much of the detection skill comes from the satellite observations.

In contrast, SATFRONT is trained on operational analyses that contain cold, warm, and occluded fronts and detects all three from satellite observations alone, so it does not depend on reanalysis or NWP variables and can be used in time-critical settings.

\section{Method}
\label{sec:method}

\subsection{Training data}
\label{sec:data}

\paragraph{Satellite observations.}
Our satellite observations consist of eight infrared channels from the GEO-Ring \citep{EUMETSAT2024-georing}, a dataset co-produced by EUMETSAT and the US National Oceanic and Atmospheric Administration (NOAA) that combines observations from the Meteosat, GOES, and Himawari geostationary satellites into a single global product with a spatial resolution of approximately 5\,km and a new image every 30 minutes. These channels are common to the satellites covering our domain and have central wavelengths of 3.8, 6.2, 7.3, 8.6, 9.7, 11.0, 12.0, and 13.3\,$\mu$m. Each channel measures the infrared radiation emitted by clouds and the Earth's surface within a narrow wavelength band, expressed as a brightness temperature, i.e., the temperature of a black body that would emit the same amount of radiation. We use only infrared channels, and not solar channels, because they are available during both day and night. GEO-Ring is provided on a HEALPix grid \citep{Gorski2005-healpix}, whose cells have equal area. Because our model requires its inputs on a regular two-dimensional grid, and ERA5 and the frontal labels are given in latitude and longitude, we map the observations to a regular $0.05^\circ$ latitude--longitude grid over Europe and the Northeast Atlantic using nearest-neighbour interpolation. The grid cells are about 5.6\,km wide in the north--south direction and, depending on latitude, between about 1 and 5\,km wide in the east--west direction, so at higher latitudes nearest-neighbour interpolation repeats values in neighbouring grid cells. The model is trained and evaluated on the same grid, but we did not test whether an equal-area grid would change its performance.

\begin{figure*}[t]
\centering
\includegraphics[width=0.8\linewidth]{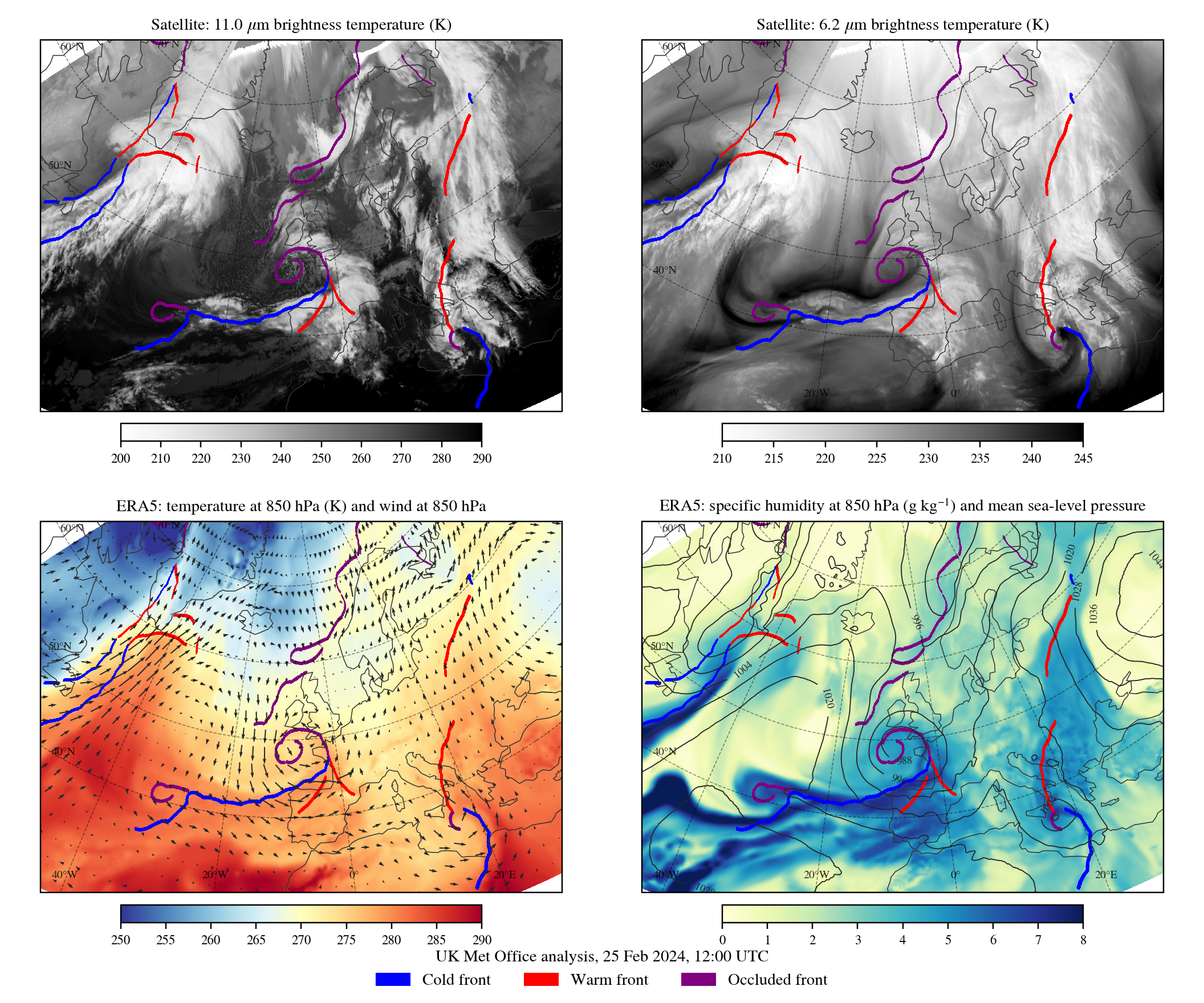}
\caption{Satellite observations (top) and ERA5 variables (bottom) for 25 February 2024, 12:00\,UTC, with the fronts of the UK Met Office analysis overlaid. Top left: the 11.0\,$\mu$m brightness temperature is close to the temperature of cloud tops, so high, cold clouds appear bright. Top right: the 6.2\,$\mu$m brightness temperature mainly reflects water vapour in the upper troposphere. Bottom left: ERA5 temperature and wind at 850\,hPa, about 1.5\,km above sea level. Bottom right: ERA5 specific humidity at 850\,hPa with mean sea level pressure contours (hPa). The satellite channels show clouds and moisture, whereas ERA5 provides temperature, humidity, wind, and pressure directly.}
\label{fig:inputs}
\end{figure*}

\paragraph{Reanalysis.}
ERA5 \citep{Hersbach2020-or} is an hourly, gridded reconstruction of past atmospheric states produced by the European Centre for Medium-Range Weather Forecasts (ECMWF). It combines historical observations with an NWP model to reconstruct the state of the atmosphere. We select 28 ERA5 variables commonly used in front detection \citep{Lagerquist2019-bg, Biard2019-to, Niebler2022-ih}: six variables (temperature, the eastward and northward wind components, specific humidity, relative humidity, and geopotential height) at four pressure levels (1000, 925, 850, and 700\,hPa), together with four surface variables (2\,m temperature, the two wind components at 10\,m, and mean sea level pressure). We take these variables from the ERA5 hourly data on pressure levels and on single levels \citep{C3S2018-era5pl, C3S2018-era5sl}, accessed through the Earth Data Hub of the Destination Earth platform. Pressure levels are surfaces of constant atmospheric pressure; the four levels used here extend from approximately sea level (1000\,hPa) to 3\,km altitude (700\,hPa). ERA5 has a spatial resolution of $0.25^\circ$, which corresponds to about 28\,km in the north--south direction. We bilinearly interpolate all ERA5 variables to the $0.05^\circ$ grid used for the satellite observations. Figure~\ref{fig:inputs} shows an example of satellite observations and ERA5 variables for the same analysis time.

\paragraph{Auxiliary features.}
Following ASII \citep{NWCSAF2022-asii}, we create seven additional features from the brightness temperatures of the 6.2, 11.0, and 12.0\,$\mu$m channels. The first two are differences between channels. The difference between the 6.2\,$\mu$m water vapour channel and the 11.0\,$\mu$m channel is close to zero for high, thick clouds and strongly negative in cloud-free regions, so it separates high frontal cloud bands from lower clouds and clear air. The difference between the 12.0 and 11.0\,$\mu$m channels depends on the optical thickness and phase of clouds and helps to separate thin from thick clouds. The remaining five are spatial filters of the 11.0\,$\mu$m channel, whose brightness temperature is close to the temperature of the cloud tops and therefore shows the structure of the clouds most directly. The filters are designed to highlight local anomalies, cloud texture, cloud edges, curved cloud structures, and large-scale cloud bands. All seven features use satellite observations only. Appendix~\ref{app:features} lists all input channels, describes each feature, and provides its formula, while Appendix~\ref{sec:aux_selection} evaluates their effect on model performance.

\paragraph{Input configurations.}
We evaluate two input configurations. The satellite-only configuration contains the eight satellite channels and the seven auxiliary features. The satellite+ERA5 configuration additionally includes the 28 ERA5 variables, so the two configurations differ only in the presence of ERA5. Because ERA5 becomes available only after a delay of several days, we use the satellite+ERA5 configuration as a diagnostic upper bound on what reanalysis adds to the model, not as an intended operational model.

\subsection{Evaluation data}

\begin{figure*}[t]
\centering
\includegraphics[width=0.85\linewidth]{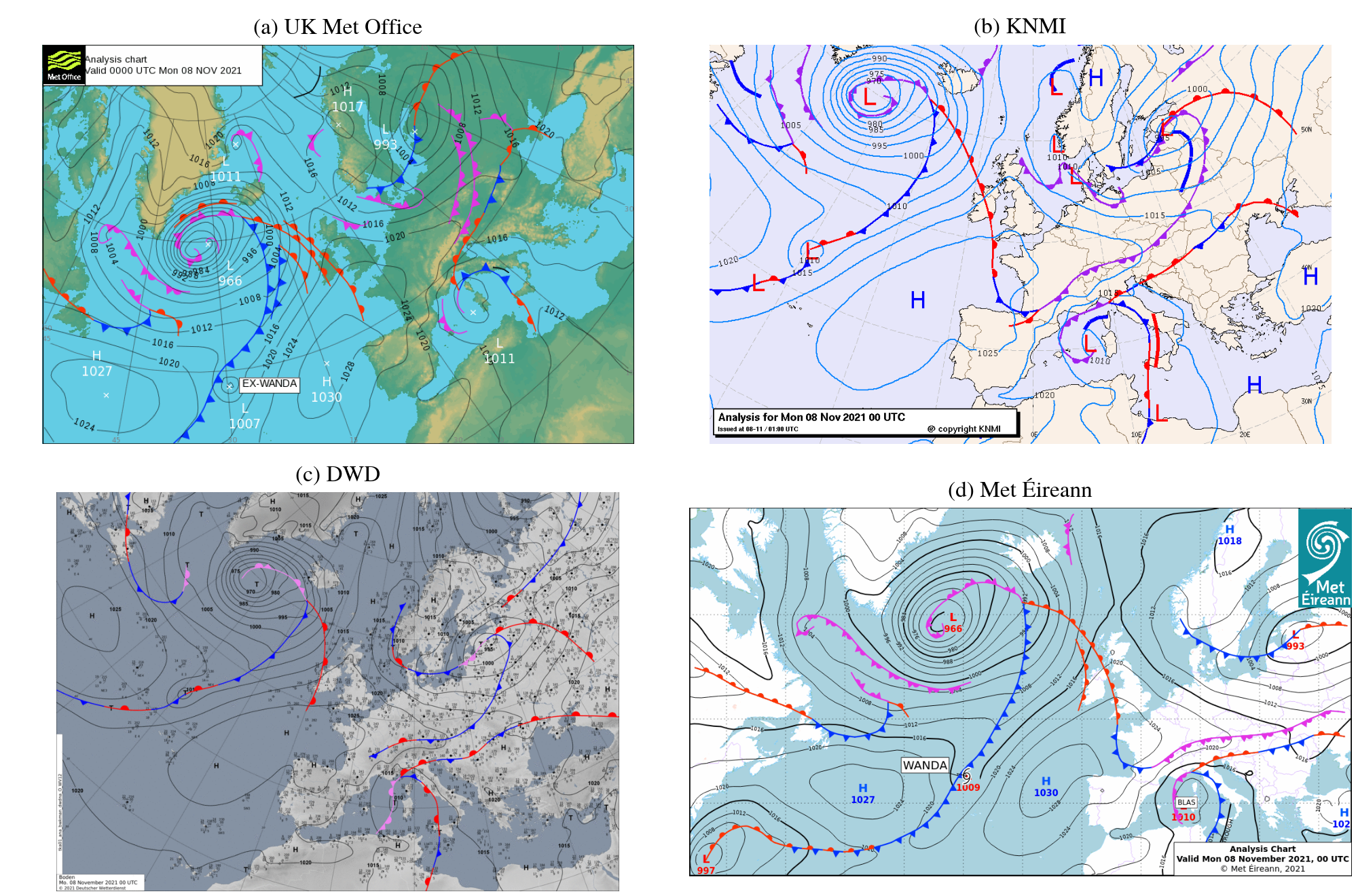}
\caption{Surface analysis charts for 8 November 2021, 00:00\,UTC, from (a) the UK Met Office, (b) KNMI, (c) DWD, and (d) Met \'Eireann. Blue lines with triangles, red lines with semicircles, and purple lines with alternating symbols mark cold, warm, and occluded fronts. The services agree on the main frontal systems but differ in the position, extent, and type of several fronts. Chart sources: (a) \textcopyright{} Crown Copyright 2021. Information provided by the National Meteorological Library and Archive -- Met Office, UK. (b) KNMI (CC0). (c) Deutscher Wetterdienst (DWD), obtained via Wetterzentrale.de. (d) \textcopyright{} Met \'Eireann, source: met.ie.}
\label{fig:charts}
\end{figure*}

Frontal labels come from operational surface analysis charts produced by five European national meteorological services: the Finnish Meteorological Institute (FMI), the German Weather Service (DWD), the Royal Netherlands Meteorological Institute (KNMI), the Irish Meteorological Service (Met \'Eireann), and the UK's national meteorological service (UK Met Office). Surface analysis charts are weather maps on which forecasters draw fronts and pressure systems (Figure~\ref{fig:charts}). We use these labels for both training and evaluation.

Analyses are available every 6 hours for FMI, KNMI, and the UK Met Office, and every 24 hours for DWD and Met \'Eireann. FMI provides the coordinates of its fronts in a machine-readable format. The other four services (DWD, KNMI, Met \'Eireann, and the UK Met Office) publish their charts only as images, from which we extract the frontal lines using a computer vision pipeline that we developed, by georeferencing each chart using manually selected control points, then isolating frontal lines based on their colour and converting them to lines in geographic coordinates. Appendix~\ref{app:extraction} describes the pipeline in more detail.

We rasterise all fronts onto the $0.05^\circ$ grid and widen each line by four pixels (approximately 20\,km) on each side. This produces segmentation masks with four classes: background, cold front, warm front, and occluded front. Each service draws fronts only within its own region of interest. We therefore define a fixed region for each service (for example, $59^\circ$W--$28^\circ$E and $38^\circ$--$76^\circ$N for the UK Met Office) and treat pixels outside that region as unlabelled rather than as background. A sample consists of the input data at a single analysis time together with the labels from one service at that time.

We use the data for the years 2019--2022 for training, 2023 for validation, and 2024 for testing. The test set contains 1\,462 of the 1\,464 six-hourly times in 2024. We exclude 4 January, 18:00\,UTC, for which our GEO-Ring dataset contains no observations, and 30 December, 18:00\,UTC, for which the archived UK Met Office chart file is not a valid image. We select between model configurations using the overall CSI at a 150\,km tolerance on the validation set (Section~\ref{sec:training_eval}).

\subsection{Architecture}
\label{sec:architecture}

\begin{figure*}[t]
\centering
\includegraphics[width=\linewidth]{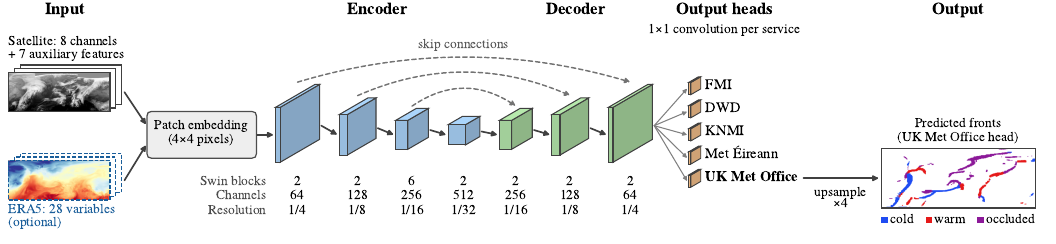}
\caption{The SATFRONT architecture with service-specific output heads. Each stack of input planes represents a group of input channels. The front planes show the 11.0\,$\mu$m satellite channel and the ERA5 temperature at 850\,hPa for 25 February 2024, 12:00\,UTC. The ERA5 variables (dashed outline) are used only in the satellite+ERA5 configuration. The encoder reduces the spatial resolution in four stages of Swin Transformer blocks, and the decoder restores it and combines its features with those of the encoder through skip connections (dashed arrows). The values below the blocks give, for each stage, the number of Swin Transformer blocks, the number of channels, and the spatial resolution relative to the input. Each output head is a $1{\times}1$ convolution that maps the shared features to class scores. The output plane shows the fronts predicted by the UK Met Office head of SATFRONT with satellite+ERA5 inputs for the same time. The single-service models have one output head.}
\label{fig:architecture}
\end{figure*}

Our model is based on a Swin-UNet \citep{Cao2023-mt}, which combines the encoder--decoder structure and skip connections of U-Net \citep{Ronneberger2015-me} with Swin Transformer blocks \citep{Liu2021-ng}. Its encoder is also motivated by Aurora \citep{Bodnar2025-ze}, an atmospheric foundation model that uses hierarchical Swin Transformer processing for high-resolution, multi-channel atmospheric fields. We follow this design principle by using a patch-based, multi-stage Swin encoder with local shifted-window attention to process our satellite and atmospheric inputs. The segmentation-specific parts of the architecture, however, follow the Swin-UNet design: we use a symmetric decoder with skip connections and a final per-pixel classification layer. We do not use Aurora's pretrained weights or its complete architecture; the model is trained from scratch for front detection. We use this architecture in SATFRONT (Figure~\ref{fig:architecture}).

The encoder first divides the input into patches of $4\times4$ pixels and then reduces the spatial resolution across four stages. Each stage computes self-attention within local windows of the feature map rather than across the entire grid, reducing the computational cost for large spatial inputs. Every second block shifts the windows, allowing attention to connect pixels in neighbouring windows.

The decoder mirrors the encoder and progressively restores the spatial resolution. At each stage, it combines the upsampled features with encoder features at the corresponding resolution through U-Net-style skip connections. A final layer maps the resulting representation to class scores for each pixel. The architecture can accommodate different numbers of input channels, allowing us to use the same network for the satellite-only and satellite+ERA5 input configurations.

We selected this architecture by comparing six convolutional and transformer-based segmentation models trained with satellite+ERA5 inputs and UK Met Office labels, the labels that we use for evaluation (Section~\ref{sec:experiments}). The six models are the Baseline U-Net described in Section~\ref{sec:experiments}; a U-Net extended with residual blocks \citep{He2016-xx}, atrous spatial pyramid pooling \citep{Chen2018-wa}, and attention gates \citep{Oktay2018-attention}; TransUNet \citep{Chen2021-transunet}; SegFormer \citep{Xie2021-segformer} with its original decoder and with a U-Net decoder; and the Swin-UNet architecture used in SATFRONT, which achieved the highest validation CSI among the models considered.

\subsection{Service-specific output heads}
\label{sec:service_heads}

Training on labels from multiple services creates a trade-off between benefiting from a larger labelled dataset and mixing service-specific frontal analysis conventions. We address this by sharing the entire network across all services except for the final layer. We use a separate final layer for each service, which we call an output head (Figure~\ref{fig:architecture}).

For each training sample, the shared network and only the output head corresponding to the service that produced the label are updated. In this way, all services contribute to learning the shared representation, while each output head can specialise to the conventions of a particular service. At evaluation time, we use the head corresponding to the service whose labels are used as ground truth.

\subsection{Training and evaluation}
\label{sec:training_eval}

\paragraph{Training.}
We train the model using the sum of a class-weighted cross-entropy loss $\mathcal{L}_{\mathrm{CE}}$ and a deformable intersection over union (IoU) loss $\mathcal{L}_{\mathrm{IoU}}$ \citep{Niebler2022-ih}, with equal weights,
\begin{equation}
    \mathcal{L} = \tfrac{1}{2}\,\mathcal{L}_{\mathrm{CE}} + \tfrac{1}{2}\,\mathcal{L}_{\mathrm{IoU}}, \quad
    \mathcal{L}_{\mathrm{IoU}} = \frac{\sum_{c} w_c \left(1 - \mathrm{IoU}_c\right)}{\sum_{c} w_c},
\end{equation}
\begin{equation}
    \mathrm{IoU}_c = \frac{\sum_{u} p_c(u)\, \tilde{y}_c(u) + 1}{\sum_{u} \bigl(p_c(u) + \tilde{y}_c(u) - p_c(u)\, \tilde{y}_c(u)\bigr) + 1},
\end{equation}
where $p_c(u)$ is the predicted probability of class $c$ at pixel $u$, $\tilde{y}_c(u) \in \{0,1\}$ is the deformed target described below, and the class weights $w_c$ are 1 for background and 15, 20, and 15 for cold, warm, and occluded fronts. Before comparing the target and predicted fronts, the deformable loss locally shifts each target front towards the predicted front, reducing the penalty for small positional differences in the hand-drawn labels. To do so, it searches for a path along high values of the predicted frontal probability that lies within three pixels of the target front and uses this path, drawn three pixels wide, as the deformed target $\tilde{y}$. The cross-entropy loss uses the same class weights and deformed targets. We compute the loss only over pixels within the region of the service that produced the label.

We stop training when the validation loss has not decreased for 10 epochs and evaluate the checkpoint with the lowest validation loss.

\paragraph{Neighbourhood scores.}
The model assigns each pixel the class with the highest predicted score. Our main metric is the neighbourhood CSI \citep{Justin2025-ti}, which does not penalise positional errors smaller than a specified distance tolerance. For a given class and tolerance, a target frontal pixel is counted as a hit if the model predicts the same class within that distance. A predicted frontal pixel is counted as a false alarm if no target pixel of the same class lies within the tolerance. With $H$ hits, $T$ target pixels, and $F$ false alarms,

\begin{equation}
    \mathrm{CSI} = \frac{H}{T + F}.
\end{equation}

CSI equals 1 only when the model detects every target pixel without producing false alarms.

We also report four related metrics, the probability of detection (POD), the success ratio (SR), the F1 score, and the frequency bias (FB). With $P$ predicted pixels,
\begin{equation}
\begin{aligned}
    \mathrm{POD} &= \frac{H}{T}, &
    \mathrm{SR} &= 1 - \frac{F}{P}, \\
    \mathrm{F1} &= \frac{2\,\mathrm{POD}\cdot\mathrm{SR}}{\mathrm{POD} + \mathrm{SR}}, &
    \mathrm{FB} &= \frac{P}{T}.
\end{aligned}
\end{equation}
POD and SR are the neighbourhood equivalents of recall and precision, F1 is their harmonic mean, and FB values above 1 indicate overprediction.

We sum the underlying counts over all test samples, compute each metric separately for each class, and report the overall score as the mean over cold, warm, and occluded fronts, for example $\mathrm{CSI}_{\mathrm{overall}} = (\mathrm{CSI}_{\mathrm{cold}} + \mathrm{CSI}_{\mathrm{warm}} + \mathrm{CSI}_{\mathrm{occluded}})/3$. Following \citet{Justin2025-ti}, we evaluate tolerances from 50 to 250\,km. We report most results at 150\,km, the midpoint of this range. All scores are computed only over pixels within the region of the service providing the ground truth.

\paragraph{Annual frontal frequency.}
To evaluate the spatial distribution of predicted fronts over the test year, we compute a frequency map for each class and for all fronts combined. For each pixel, the map gives the fraction of the 1\,462 test samples, corresponding to one sample every 6 hours, in which that pixel belongs to a front. We compute these maps for both the predictions and the ground truth and compare them using three measures, all calculated on the unsmoothed maps. Unlike the five metrics above, which compare predictions and labels at individual analysis times, these measures compare the annual maps.

Let $f^{\mathrm{GT}}_i$, $f^{\mathrm{S}}_i$, and $f^{\mathrm{SE}}_i$ denote the frequencies of the ground truth and of the satellite-only and satellite+ERA5 predictions at pixel $i$ of the $N$ pixels in the region. The spatial correlation $r$ is the Pearson correlation between a predicted map and $f^{\mathrm{GT}}$ and measures the similarity of their spatial patterns independently of the overall frontal frequency. The error reduction is the relative decrease in the mean squared error (MSE) from the ground-truth map when ERA5 is added, equivalent to the Murphy skill score (MSS) \citep{Murphy1988-bm},
\begin{equation}
\begin{aligned}
    \mathrm{MSS} &= 1 - \frac{\mathrm{MSE}(f^{\mathrm{SE}})}{\mathrm{MSE}(f^{\mathrm{S}})}, \\
    \mathrm{MSE}(f) &= \frac{1}{N}\sum_{i=1}^{N} \bigl(f_i - f^{\mathrm{GT}}_i\bigr)^2.
\end{aligned}
\end{equation}
Finally, $W$ is the percentage of pixels for which the satellite+ERA5 map is strictly closer to the ground truth than the satellite-only map,
\begin{equation}
    W = \frac{100\%}{N}\sum_{i=1}^{N} \mathbf{1}\bigl[\,|f^{\mathrm{SE}}_i - f^{\mathrm{GT}}_i| < |f^{\mathrm{S}}_i - f^{\mathrm{GT}}_i|\,\bigr].
\end{equation}
where $\mathbf{1}[\cdot]$ equals 1 if the condition holds and 0 otherwise. Pixels at which both maps are equally close to the ground truth count as not closer. Such ties occur at 4--14\% of the pixels, depending on the class, so $W$ should be compared with the corresponding percentage for the satellite-only map rather than with 50\%.

\section{Experiments}
\label{sec:experiments}

Unless stated otherwise, we report results on the 2024 test set using the UK Met Office analyses as ground truth. We use the UK Met Office because its archive combines 6-hourly analyses with broad spatial coverage, and because we can convert its charts to labels with fewer extraction errors than the charts of other services. SATFRONT refers to the model trained on labels from all five services using the service-specific output heads described in Section~\ref{sec:service_heads}, evaluated with its UK Met Office head. We also report results for the same architecture trained only on UK Met Office labels, which we refer to as single-service SATFRONT.

We compare these models with two baselines. The first is a U-Net \citep{Ronneberger2015-me} trained on UK Met Office labels only, using the same inputs and loss as our model (Baseline U-Net). The second is the operational ASII product from NWC SAF, evaluated both in its satellite-only mode (ASII-SAT) and in the mode that additionally uses NWP data (ASII-SAT+NWP). We rasterise the ASII fronts onto our grid and evaluate them with the same evaluation procedure as our models, within the UK Met Office region and on the 1\,453 test samples for which ASII output is available.

\begin{figure*}[t]
\centering
\includegraphics[width=\linewidth]{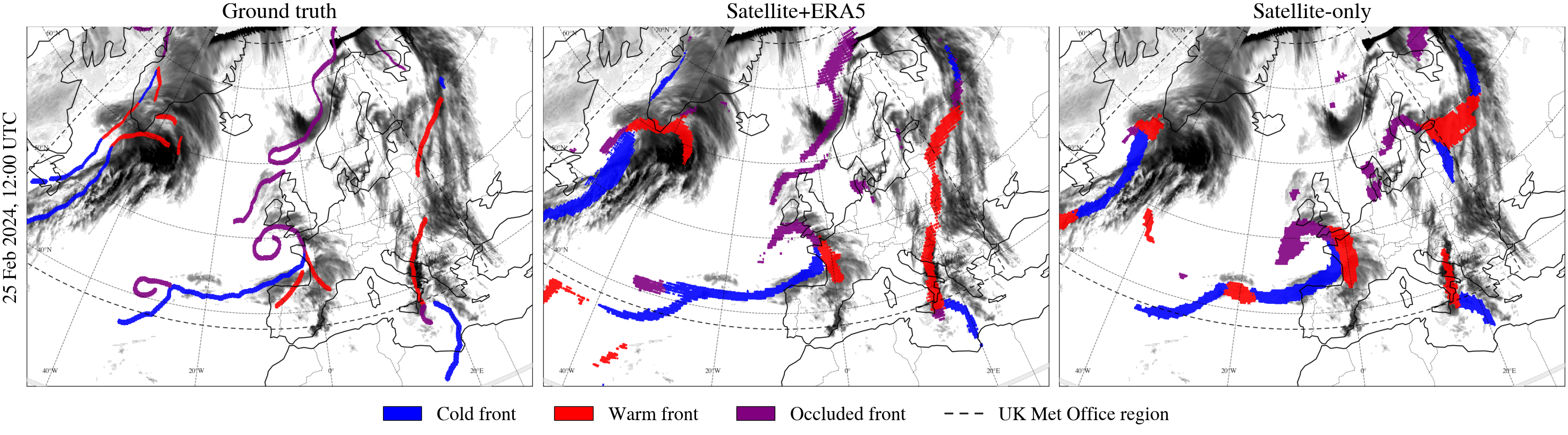}
\caption{Ground truth (left) and predictions of SATFRONT with satellite+ERA5 inputs (middle) and satellite-only inputs (right) for 25 February 2024, 12:00\,UTC, using the UK Met Office head. The grey background shows the 11.0\,$\mu$m brightness temperature where it is below 250\,K, corresponding to cold cloud tops. The dashed line marks the UK Met Office region (59$^{\circ}$W--28$^{\circ}$E, 38$^{\circ}$--76$^{\circ}$N). Only pixels inside this region are scored, and only UK Met Office labels inside it contribute to the training loss. Ground-truth fronts that extend beyond the region are therefore not used. For visualisation, the satellite-only predictions are shown after a post-processing step that fills gaps of up to approximately 12 pixels using a morphological closing and removes regions smaller than 120 pixels. The satellite+ERA5 predictions are shown without post-processing. All quantitative results in this paper use predictions without post-processing. Appendix~\ref{app:postprocessing} compares satellite-only predictions with and without post-processing.}
\label{fig:examples}
\end{figure*}

\begin{figure}[t]
\centering
\includegraphics[width=\columnwidth]{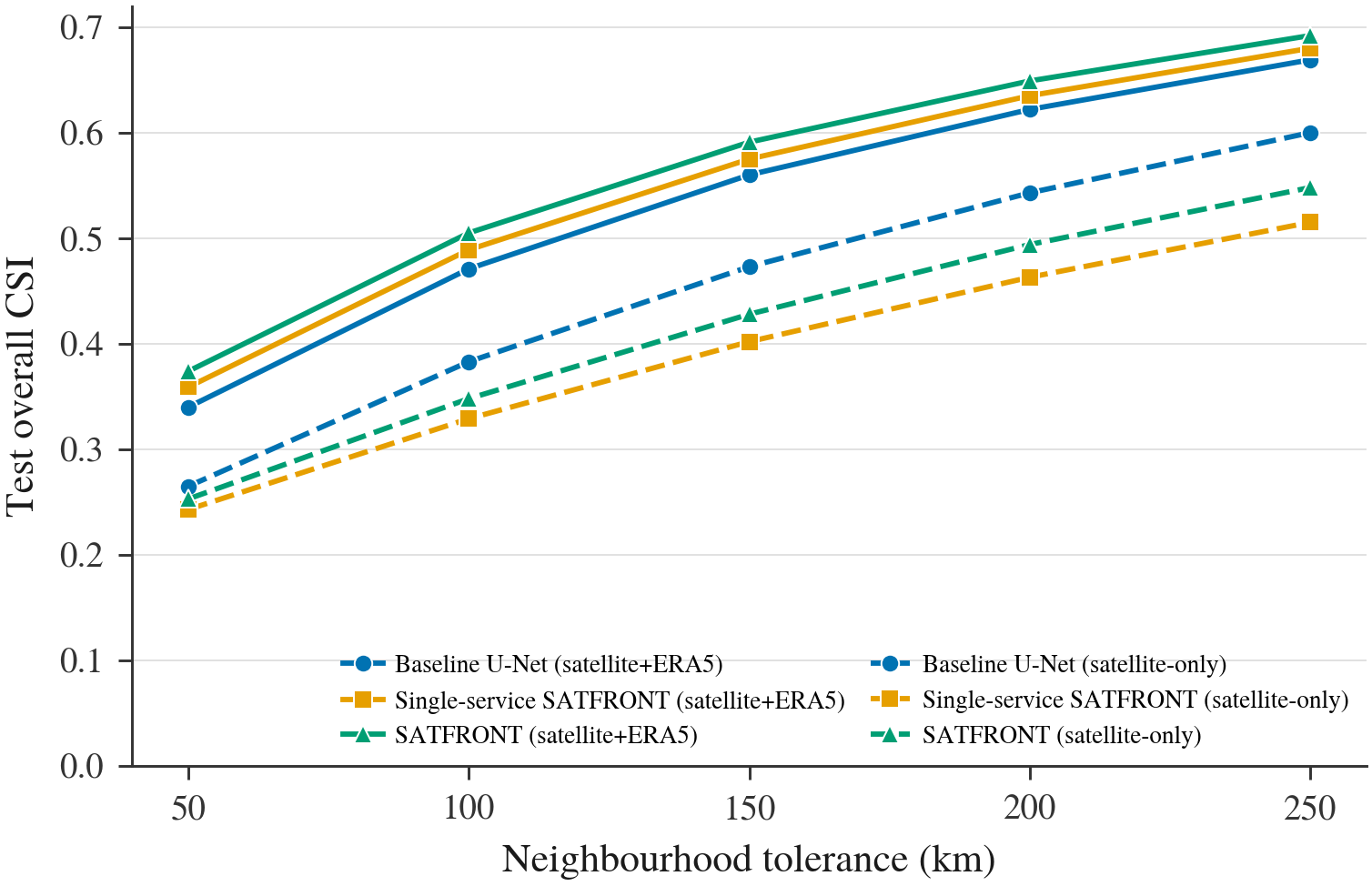}
\caption{Overall CSI on the 2024 test set as a function of the tolerance (50--250\,km). Solid lines show satellite+ERA5 inputs, and dashed lines show satellite-only inputs. The Baseline U-Net and single-service SATFRONT are trained only on UK Met Office labels, whereas SATFRONT is trained on labels from all five services.}
\label{fig:perf_diagram}
\end{figure}

\subsection{Main results}
\label{sec:main_results}

Table~\ref{tab:main_results} reports the test scores of SATFRONT and the baselines.

\begin{table}[t]
\centering
\caption{Test scores at the 150\,km tolerance on the 2024 test set, with UK Met Office labels as ground truth. SATFRONT is trained on labels from all five services and evaluated with its UK Met Office head. The Baseline U-Net is trained only on UK Met Office labels. ASII-SAT and ASII-SAT+NWP are the two modes of the operational ASII product. The models are grouped by input, and bold marks the best overall value of each metric within each group (for FB, the value closest to 1).}
\label{tab:main_results}
\scriptsize
\setlength{\tabcolsep}{2.5pt}
\begin{tabular}{@{}llcccccc@{}}
\toprule
\textbf{Model} & \textbf{Input} & \textbf{Class} & \textbf{CSI} & \textbf{POD} & \textbf{SR} & \textbf{F1} & \textbf{FB} \\
\midrule
\multirow{4}{*}{SATFRONT (ours)}
    & \multirow{4}{*}{Satellite}
    & Overall  & .428 & .561 & \textbf{.691} & .618 & \textbf{1.005} \\
    && Cold     & .462 & .600 & .734 & .660 & 1.122 \\
    && Warm     & .378 & .546 & .612 & .577 & 1.149 \\
    && Occluded & .445 & .535 & .726 & .616 & 0.743 \\
Baseline U-Net & Satellite
    & Overall  & \textbf{.473} & \textbf{.660} & .610 & \textbf{.631} & 1.025 \\
ASII-SAT & Satellite
    & Overall  & .133 & .355 & .237 & .265 & 2.332 \\
\midrule
\multirow{4}{*}{SATFRONT (ours)}
    & \multirow{4}{*}{Satellite+ERA5}
    & Overall  & \textbf{.591} & .733 & \textbf{.809} & \textbf{.769} & \textbf{1.268} \\
    && Cold     & .616 & .783 & .823 & .803 & 1.539 \\
    && Warm     & .568 & .712 & .801 & .754 & 1.270 \\
    && Occluded & .588 & .702 & .804 & .750 & 0.994 \\
Baseline U-Net & Satellite+ERA5
    & Overall  & .560 & \textbf{.769} & .763 & .765 & 1.580 \\
ASII-SAT+NWP & Satellite+NWP
    & Overall  & .143 & .331 & .240 & .262 & 1.900 \\
\bottomrule
\end{tabular}
\end{table}

\begin{figure*}[t]
\centering
\includegraphics[width=\linewidth]{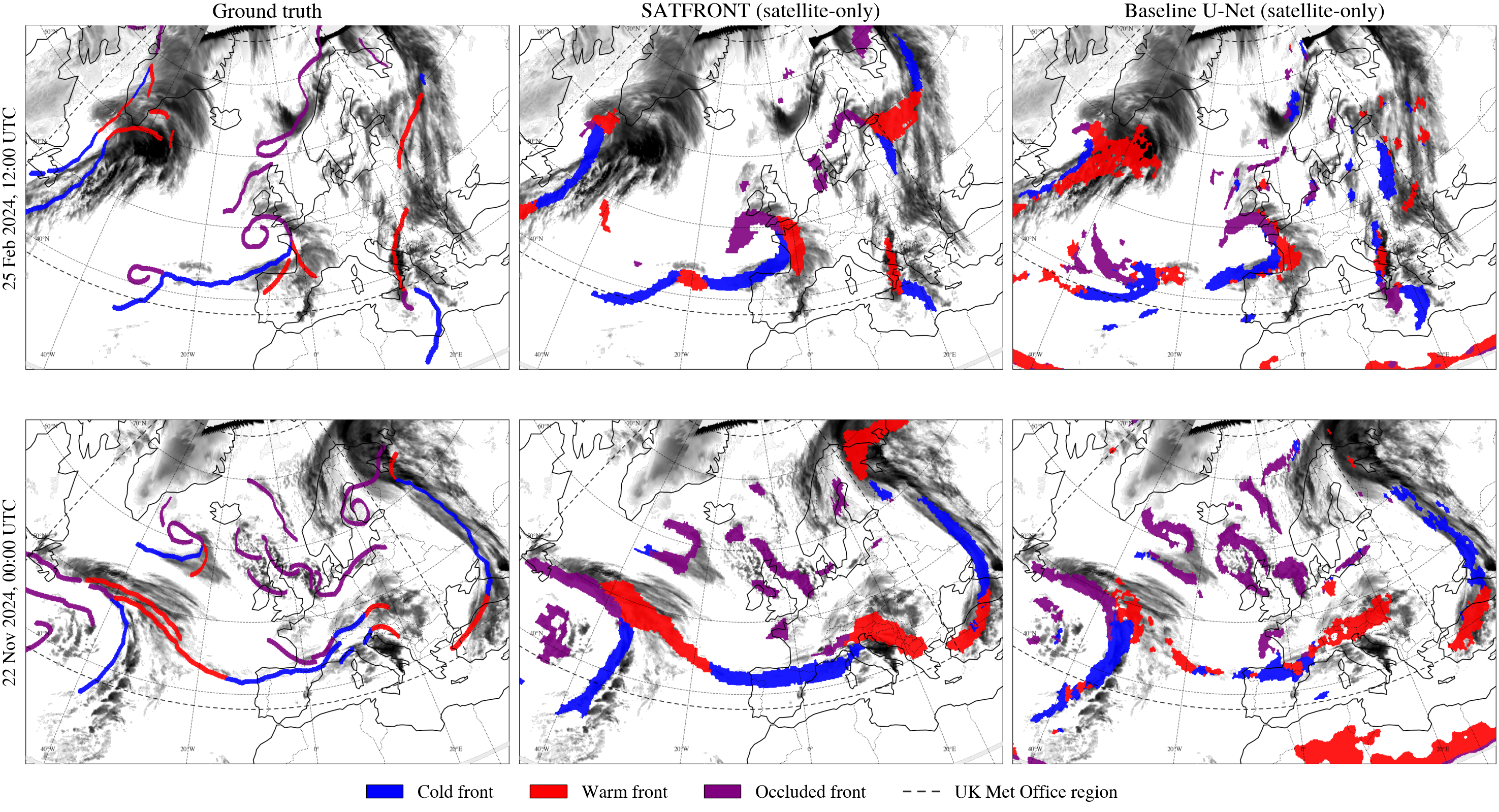}
\caption{Ground truth (left) and satellite-only predictions of SATFRONT (middle) and the Baseline U-Net (right) for 25 February 2024, 12:00\,UTC (top) and 22 November 2024, 00:00\,UTC (bottom). Predictions from both models are shown after the same post-processing used in Figure~\ref{fig:examples}, which fills small gaps and removes regions smaller than 120 pixels. The dashed line marks the UK Met Office region, as in Figure~\ref{fig:examples}. Fronts predicted outside it, for example over North Africa, are not included in the scores.}
\label{fig:unet_vs_main}
\end{figure*}

\begin{figure*}[t]
\centering
\includegraphics[width=\linewidth]{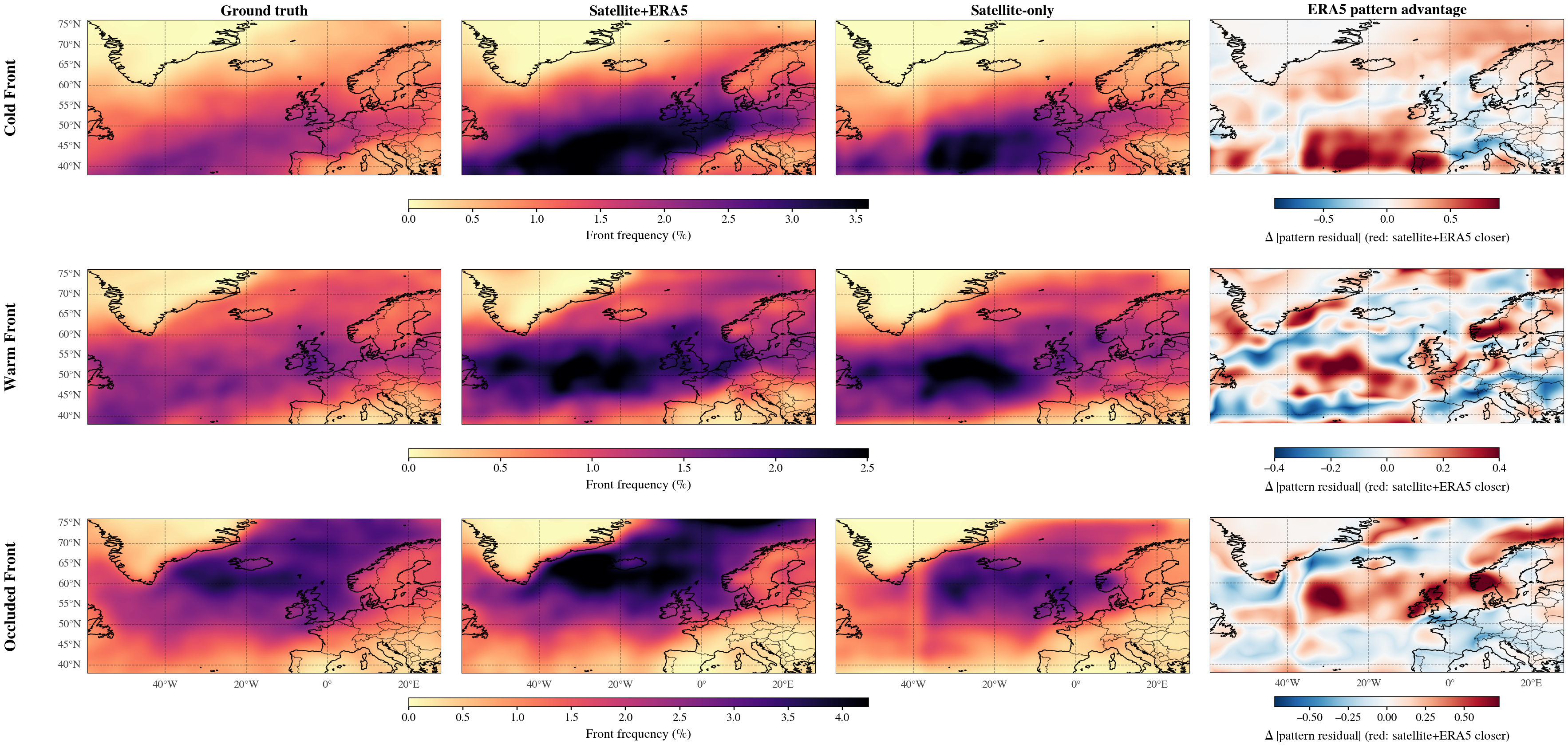}
\caption{Frequency of cold, warm, and occluded fronts (rows) over the 2024 test year within the UK Met Office region, expressed as a percentage of the 1\,462 test samples. Columns 1--3 show the ground truth and SATFRONT with satellite+ERA5 and satellite-only inputs. Column 4 (``pattern residual'') shows pixels where the satellite+ERA5 map is closer to the ground truth than the satellite-only map (red) or farther from it (blue). For this column only, each map is first divided by its mean so that the comparison reflects spatial pattern rather than overall frontal frequency. It can therefore differ from the pixel percentages reported in Table~\ref{tab:climatology}, which are computed from the unnormalised maps. All maps are smoothed for visualisation using a Gaussian filter with a standard deviation of 150\,km.}
\label{fig:climatology}
\end{figure*}

\paragraph{Overall scores.}
Using satellite observations alone, SATFRONT achieves an overall CSI of 0.428, approximately 3 times the CSI of either ASII modes (0.133 and 0.143). Adding ERA5 increases the overall CSI to 0.591 and improves both POD and SR for every front class, meaning that the satellite+ERA5 model detects a larger fraction of the fronts in the UK Met Office analysis while also producing fewer false alarms. The satellite+ERA5 model, however, overpredicts fronts (FB 1.268), mainly cold fronts (FB 1.539).

\paragraph{Scores per front class.}
For both input configurations, CSI is highest for cold fronts and lowest for warm fronts, which agrees with FrontFinder~AI results over North America \citep{Justin2025-ti}. Adding ERA5 produces the largest improvement for warm fronts, whose CSI increases from 0.378 to 0.568. One possible explanation is that warm fronts are often associated with broad and diffuse cloud structures, which can make their position and type more difficult to infer from satellite observations alone.

\paragraph{Example predictions.}
Figure~\ref{fig:examples} shows the predictions of both input configurations for one test sample. The satellite+ERA5 model places most predicted fronts close to those in the UK Met Office analysis. The satellite-only model misses some warm and occluded fronts and assigns the wrong class to a few frontal segments. Nevertheless, most of the predicted fronts are located close to a front drawn by UK Met Office forecasters.

\paragraph{Comparison with the Baseline U-Net.}
We selected the architecture based on its validation CSI with satellite+ERA5 inputs (Section~\ref{sec:architecture}). With satellite-only inputs, however, the Baseline U-Net achieves a higher CSI than both SATFRONT and single-service SATFRONT at every tolerance shown in Figure~\ref{fig:perf_diagram}. At 150\,km, the Baseline U-Net achieves a CSI of 0.473, compared with 0.402 for single-service SATFRONT, which is trained on the same labels, and 0.428 for SATFRONT. The Baseline U-Net detects a larger fraction of the target frontal pixels than SATFRONT (POD 0.660 vs.\ 0.561), but its predictions contain more false alarms, as reflected by its lower SR (0.610 vs.\ 0.691). This difference is particularly large for warm fronts (SR 0.505 vs.\ 0.612). With satellite+ERA5 inputs, the ordering is reversed: the Baseline U-Net achieves a CSI of 0.560, compared with 0.575 for single-service SATFRONT and 0.591 for SATFRONT.

\paragraph{Continuity of the predicted fronts.}
Figure~\ref{fig:unet_vs_main} illustrates an important difference that is not fully captured by CSI. The satellite-only Baseline U-Net often produces many small, disconnected frontal segments, changes class along what appears to be a single front, and predicts spurious fronts over North Africa. SATFRONT generally produces fewer and more continuous frontal structures that more closely resemble the fronts drawn by UK Met Office forecasters.

We quantify this difference using an object-level evaluation. Following \citet{Qin2024-detr}, we group predicted and target pixels of each class into connected regions and match each predicted region to at most one target region, and vice versa, when their centres are less than 5$^{\circ}$ apart. As a result, a front that is split into several predicted segments produces additional unmatched regions. Averaged over the three front classes, 95.4\% of the Baseline U-Net's predicted regions have no matching target region, compared with 91.2\% for SATFRONT. For single-service SATFRONT, the corresponding value is 94.6\%, suggesting that part of the improvement in continuity comes from training on labels from all five services.

For use as a first-guess analysis, we expect more continuous fronts with fewer false alarms to be more useful to forecasters than a larger number of fragmented predictions. For this reason, and to ensure that the satellite-only and satellite+ERA5 configurations differ only in their inputs, we use the SATFRONT architecture for both input configurations.

\paragraph{Dependence on the tolerance.}
For every model and tolerance shown in Figure~\ref{fig:perf_diagram}, the satellite+ERA5 configuration achieves a higher CSI than the corresponding satellite-only configuration. The improvement from adding ERA5 therefore does not depend on the particular tolerance. For SATFRONT with satellite+ERA5 inputs, CSI increases from 0.374 at a tolerance of 50\,km to 0.692 at 250\,km.

\paragraph{Effect of multi-service training.}
Training on labels from all five services increases the test CSI by 0.026 with satellite-only inputs (0.428 for SATFRONT vs.\ 0.402 for single-service SATFRONT) and by 0.016 with satellite+ERA5 inputs (0.591 vs.\ 0.575). Because each model was trained only once, we cannot determine whether differences of this magnitude are larger than the variation that would arise across repeated training runs.

\subsection{Agreement between services}
\label{sec:inter_agency}

To provide a reference for the CSI values achieved by the model, we replace the model predictions with the analyses of another meteorological service and evaluate them against the UK Met Office analyses using the same metric. For each service, we use only the analysis times available from both services and only the geographic region covered by both. We estimate 95\% confidence intervals using a block bootstrap over samples, with blocks of 20 consecutive samples to account for the correlation between analyses close together in time. Table~\ref{tab:inter_agency} reports the results.

\begin{table}[t]
\centering
\caption{Agreement between services on the 2024 test set, with the UK Met Office (MO) analyses as ground truth. Values are CSI at 150\,km, with 95\% block-bootstrap confidence intervals for the overall service scores. Each service row uses only the samples and geographic region shared with the UK Met Office. The SATFRONT rows use all test samples and the full UK Met Office region.}
\label{tab:inter_agency}
\scriptsize
\setlength{\tabcolsep}{3pt}
\begin{tabular}{@{}lcccc@{}}
\toprule
\textbf{Predictor} & \textbf{Overall} & \textbf{Cold} & \textbf{Warm} & \textbf{Occl.} \\
\midrule
DWD $\to$ MO           & .376\,[.363,\,.392] & .481 & .399 & .247 \\
FMI $\to$ MO           & .277\,[.263,\,.291] & .322 & .266 & .244 \\
KNMI $\to$ MO          & .468\,[.452,\,.480] & .522 & .436 & .446 \\
Met \'Eireann $\to$ MO & .501\,[.486,\,.519] & .543 & .454 & .508 \\
\midrule
SATFRONT, satellite-only $\to$ MO & .428 & .462 & .378 & .445 \\
SATFRONT, satellite+ERA5 $\to$ MO & .591 & .616 & .568 & .588 \\
\bottomrule
\end{tabular}
\end{table}

With satellite observations alone, SATFRONT achieves a CSI of 0.428. This lies between the DWD value of 0.376 and the KNMI value of 0.468, and therefore within the range of CSI values obtained when the analyses of the other services are compared with those of the UK Met Office. With satellite+ERA5 inputs, SATFRONT achieves 0.591, which is 0.090 higher than the value for Met \'Eireann (0.501), the service whose analyses have the highest CSI against the UK Met Office in this comparison.

This comparison has three caveats.
\begin{itemize}
    \item CSI between two meteorological services measures agreement between their analyses, not the quality of either analysis. Different services may place, connect, or classify the same front differently because they follow different conventions, use different local observations, and focus on different geographic regions. The inter-service CSI values should therefore be interpreted as a reference for the degree of agreement between independently produced analyses, rather than as an upper bound on achievable model performance or the quality of the labelling.
    \item The UK Met Office output head of our model is trained directly on UK Met Office labels and can therefore learn the conventions of that service. The other meteorological services are independent and do not attempt to reproduce UK Met Office conventions.
    \item The service-to-service scores are computed only over the analysis times and region that each service shares with the UK Met Office, whereas the model scores use all test samples and the full UK Met Office region. The model scores and the service-to-service scores are therefore not directly comparable.
\end{itemize}

\subsection{Comparison with the operational baseline}
\label{sec:asii}

The ASII product from NWC SAF achieves an overall CSI of 0.133 in its satellite-only mode and 0.143 when NWP information is also included (Table~\ref{tab:main_results}). SATFRONT with satellite-only inputs achieves 0.428, compared with 0.133 for ASII-SAT, which also uses satellite observations only. With satellite+ERA5 inputs, SATFRONT achieves 0.591, compared with 0.143 for ASII-SAT+NWP. The largest difference is in SR. ASII-SAT has an SR of 0.237, compared with 0.691 for satellite-only SATFRONT, so most of its predicted frontal pixels have no front of the same class in the UK Met Office analysis within 150\,km. ASII-SAT also predicts more than twice as many frontal pixels as are present in the UK Met Office ground truth (FB 2.332, compared with 1.005 for satellite-only SATFRONT) and detects a smaller fraction of the fronts in the UK Met Office analysis (POD 0.355, compared with 0.561).

The CSI difference between ASII and our models does not reflect only the difference between manually designed rules and a model learned from data. ASII is not trained on the frontal analyses of any meteorological service, so its rules do not necessarily match the conventions of the UK Met Office or any other individual service. The UK Met Office head of SATFRONT, in contrast, is trained on the same labels against which it is evaluated. Part of the difference in CSI is therefore likely to reflect differences in frontal analysis conventions.

\subsection{Annual frontal frequency}
\label{sec:climatology}

CSI evaluates how well fronts are placed in individual analysis times. We additionally evaluate whether the models reproduce the spatial distribution of frontal activity over the test year using the frequency maps and measures defined in Section~\ref{sec:training_eval}. Figure~\ref{fig:climatology} shows the maps, and Table~\ref{tab:climatology} reports the corresponding quantitative comparisons.

Both model configurations reproduce the main large-scale pattern in the ground-truth frequency maps: frontal activity is highest over the North Atlantic and lower over southern Europe and the Mediterranean. Adding ERA5 increases the spatial correlation with the ground truth for every front class. For all fronts combined, for example, the correlation increases from 0.402 with satellite-only inputs to 0.559 with satellite+ERA5 inputs.

The other two measures reveal a more class-dependent picture. For cold and occluded fronts, the satellite+ERA5 frequency maps are not consistently closer to the ground truth than the satellite-only maps. Their mean squared errors are slightly larger, by 3.5\% for cold fronts and 2.0\% for occluded fronts, and the satellite+ERA5 map is closer to the ground truth at 43.0\% and 50.1\% of pixels, respectively, compared with 42.9\% and 42.1\% for the satellite-only map.

For cold fronts, this may initially appear inconsistent with the higher spatial correlation of the satellite+ERA5 map. The two metrics, however, measure different properties. The correlation describes whether areas of relatively high and low frontal frequency occur in similar locations, but is insensitive to a uniform difference in the overall number of predicted fronts. The mean squared error  on the other hand is sensitive to both spatial pattern and magnitude. The satellite+ERA5 model overpredicts cold-front pixels (FB 1.539 in Table~\ref{tab:main_results}), so its cold-front frequency map can have a more accurate spatial pattern while still having a larger absolute error. For warm fronts, in contrast, adding ERA5 reduces the mean squared error by 48.5\%.

These results suggest that ERA5 contributes differently depending on the front type and evaluation scale. For cold and occluded fronts, adding ERA5 substantially improves front placement in individual samples, increasing CSI by 0.154 and 0.143, respectively, while producing annual frequency map errors similar to those of the satellite-only model. For warm fronts, ERA5 improves both the placement of individual fronts and their annual spatial distribution.

\begin{table}[t]
\centering
\caption{Comparison of the annual frontal frequency maps for the 2024 test year with the UK Met Office ground truth, using the measures defined in Section~\ref{sec:training_eval}. A negative error reduction indicates that the satellite+ERA5 map has a larger mean squared error than the satellite-only map.}
\label{tab:climatology}
\scriptsize
\setlength{\tabcolsep}{3pt}
\begin{tabular}{@{}lcccc@{}}
\toprule
& \multicolumn{2}{c}{\textbf{Correlation} $r$} & \textbf{Error} & \textbf{Satellite+ERA5} \\
\textbf{Class} & Satellite-only & Satellite+ERA5 & \textbf{reduction} (MSS) & \textbf{closer} ($W$) \\
\midrule
All fronts & .402 & .559 & $+$20.5\% & 54.1\% \\
Cold       & .649 & .764 & $-$3.5\% & 43.0\% \\
Warm       & .345 & .521 & $+$48.5\% & 53.6\% \\
Occluded   & .420 & .550 & $-$2.0\% & 50.1\% \\
\bottomrule
\end{tabular}
\end{table}

\section{Limitations}
\label{sec:limitations}

We train SATFRONT on four years (2019--2022) of data over Europe and the Northeast Atlantic, which cannot capture the full range of weather situations or establish how well the model generalises to other regions. Although GEO-Ring covers most of the globe outside the polar regions, we evaluate the model only against UK Met Office analyses and only within the UK Met Office region.

The labels from four of the five services (DWD, KNMI, Met \'Eireann, and the UK Met Office) are extracted automatically from chart images and may therefore contain extraction errors. We nevertheless use the UK Met Office analyses as ground truth because they combine 6-hourly analyses with broad spatial coverage and can be converted to labels with fewer extraction errors than the charts of some other services.

In addition, each model configuration is trained only once, because a single training run of SATFRONT takes between 64 and 98 hours on one GPU. We therefore cannot determine whether small performance differences between models exceed the variability between training runs, and we do not estimate confidence intervals for these model-to-model differences.

Because ERA5 becomes available only after several days, the satellite+ERA5 configuration cannot be used in time-critical applications. Replacing ERA5 with operational NWP output would not remove this limitation, because NWP analyses also become available only a few hours after the time they describe and forecasts contain no observations made after their initialisation time. In addition, the model would probably need to be retrained or fine-tuned on the NWP output, which we have not tested.

Finally, we do not correct the satellite observations for parallax, the apparent displacement of high clouds away from the satellite when viewed at an oblique angle. This displacement can reach tens of kilometres and may therefore contribute to errors in the predicted front locations.

\section{Conclusion}
\label{sec:conclusion}

We introduced SATFRONT, to our knowledge the first deep learning framework that detects cold, warm, and occluded fronts from geostationary satellite observations alone. Evaluated against the 2024 UK Met Office analyses, our satellite-only model achieves a CSI comparable to the agreement observed between independently produced analyses from different meteorological services, while substantially outperforming the operational ASII satellite product from NWC SAF. Training with service-specific output heads also allows the model to benefit from labels from multiple meteorological services while retaining their individual analysis conventions and produces more coherent frontal structures than training on UK Met Office labels alone.

Adding ERA5 substantially improves the detection and placement of individual fronts, with the largest benefit for warm fronts. The annual frontal frequency analysis provides a complementary picture, showing that ERA5 consistently improves the spatial pattern of frontal activity, while its effect on absolute frequency errors varies by front type. In particular, the satellite observations alone already capture much of the large-scale distribution of cold and occluded frontal activity, whereas warm fronts benefit more clearly from the additional atmospheric information. Together, these results show both what can be inferred directly from satellite observations and what additional information gridded atmospheric fields contribute.

Most importantly, detecting fronts from geostationary satellite observations enables running front detection as soon as new observations are available. With GEO-Ring observations available every 30 minutes, SATFRONT can produce frontal analyses at a much higher temporal frequency than the manually produced charts used for training. Running detection on geostationary satellite observations enables automated first-guess analyses for operational meteorologists. It could also help build consistent records of frontal activity in regions where machine-readable analyses are unavailable, particularly because GEO-Ring will be produced with global coverage from 1979 onwards, using the channels available at each time. Future work should evaluate the approach with additional meteorological services and geographic regions, quantify variability across repeated training runs, and ask operational forecasters to evaluate how useful the predicted fronts are as first-guess analyses.

\section*{Acknowledgements}
We thank the Finnish Meteorological Institute (FMI) for providing its frontal analyses. We also acknowledge the German Weather Service (DWD), the Royal Netherlands Meteorological Institute (KNMI), the Irish Meteorological Service (Met \'Eireann), and the UK's national meteorological service (UK Met Office), whose surface-analysis charts we used in this work. We acknowledge EUMETSAT, Destination Earth, and the European Weather Cloud for providing compute and storage resources. The authors declare no competing interests.

\bibliography{references}
\bibliographystyle{icml2026}

\appendix
\raggedbottom
\renewcommand{\dbltopfraction}{0.95}
\renewcommand{\dblfloatpagefraction}{0.95}
\renewcommand{\textfraction}{0.05}
\setcounter{dbltopnumber}{3}
\begin{table*}[b]
\centering
\caption{Input channels. The satellite-only configuration uses the 15 channels of the first two groups, and the satellite+ERA5 configuration uses all 43 channels. The descriptions of the satellite channels give their main sensitivity.}
\label{tab:inputs}
\footnotesize
\begin{tabular}{@{}llp{0.52\linewidth}@{}}
\toprule
\textbf{Group} & \textbf{Channel or variable} & \textbf{Description} \\
\midrule
Satellite (GEO-Ring, 8) & 3.8\,$\mu$m & Infrared window; low clouds and fog at night; includes reflected sunlight during the day \\
 & 6.2\,$\mu$m & Water vapour in the upper troposphere \\
 & 7.3\,$\mu$m & Water vapour in the middle troposphere \\
 & 8.6\,$\mu$m & Infrared window; cloud phase \\
 & 9.7\,$\mu$m & Ozone absorption \\
 & 11.0\,$\mu$m & Infrared window; temperature of cloud tops and the surface \\
 & 12.0\,$\mu$m & Infrared window with weak water vapour absorption \\
 & 13.3\,$\mu$m & Carbon dioxide absorption; cloud-top height \\
\midrule
Auxiliary features (7) & $F_{\mathrm{WV}}$, $F_{\mathrm{SW}}$ & Channel differences (see the list above) \\
 & $F_{\mathrm{dev}}$, $F_{\sigma}$, $F_{\nabla}$, $F_{\mathrm{H}}$, $F_{\mathrm{L}}$ & Spatial filters of the 11.0\,$\mu$m channel (see the list above) \\
\midrule
ERA5, pressure levels (24) & $T$, $u$, $v$, $q$, $RH$, $z$ & Temperature, eastward and northward wind, specific and relative humidity, and geopotential height, each at 1000, 925, 850, and 700\,hPa \\
ERA5, surface (4) & $T_{2\mathrm{m}}$, $u_{10\mathrm{m}}$, $v_{10\mathrm{m}}$, MSLP & 2\,m temperature, 10\,m wind components, and mean sea level pressure \\
\bottomrule
\end{tabular}
\end{table*}

\begin{figure*}[b]
\centering
\includegraphics[width=0.7\linewidth]{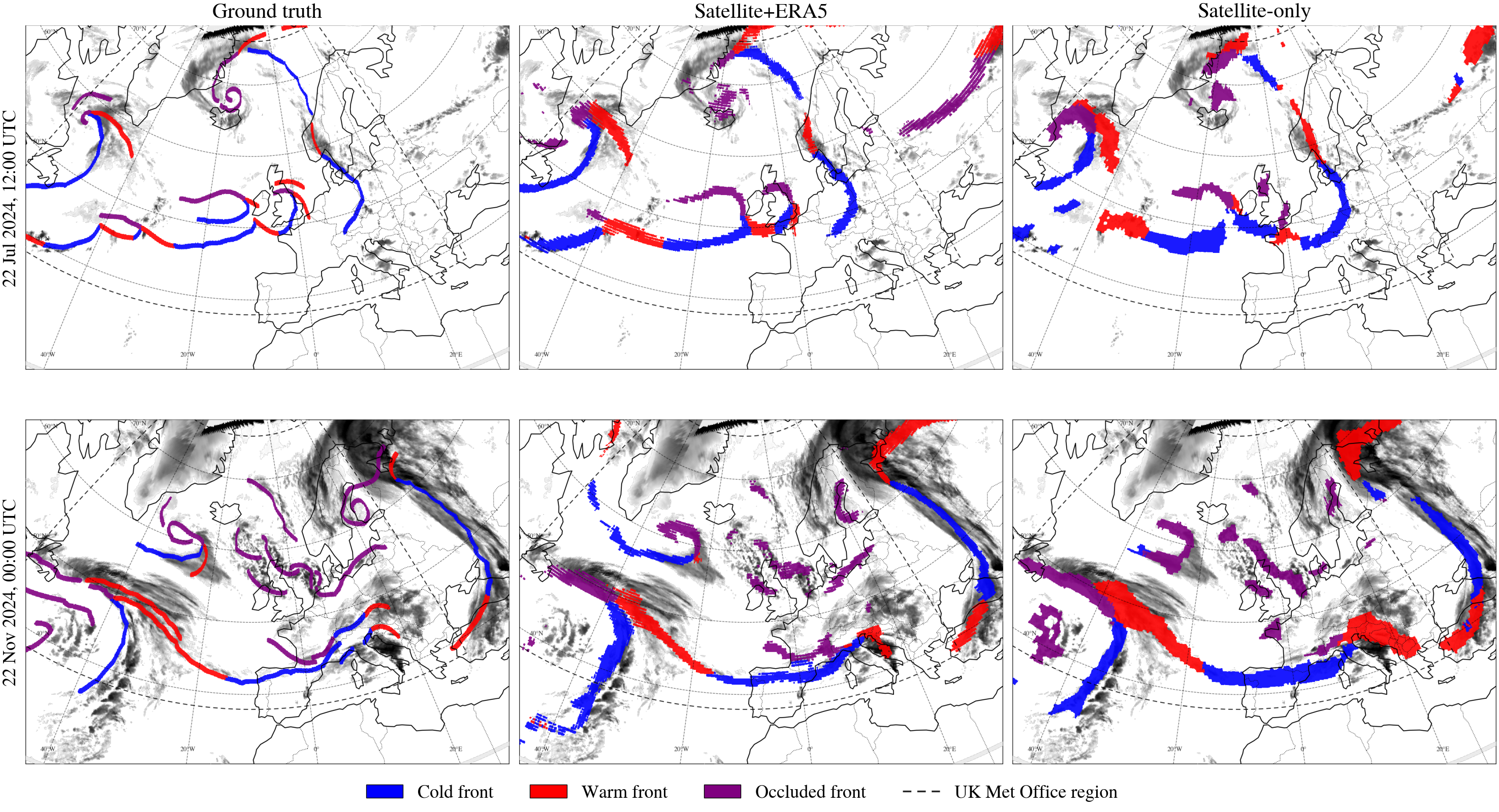}
\caption{Ground truth (left) and predictions of SATFRONT with satellite+ERA5 inputs (middle) and satellite-only inputs (right) for 22 July 2024, 12:00\,UTC (top) and 22 November 2024, 00:00\,UTC (bottom), using the UK Met Office head. As in Figure~\ref{fig:examples}, the dashed line marks the UK Met Office region, the satellite-only predictions are shown after post-processing, and the satellite+ERA5 predictions are shown without post-processing.}
\label{fig:examples_appendix}
\end{figure*}

\section{Auxiliary features}
\label{app:features}

We compute the seven auxiliary features from the brightness temperatures $T_{6.2}$, $T_{11.0}$ and $T_{12.0}$ of the 6.2, 11.0 and 12.0\,$\mu$m channels. For a pixel $p$, $\mu_5(X)(p)$ denotes the mean of $X$ over the $5{\times}5$ pixel neighbourhood of $p$.
\begin{itemize}
    \item \textbf{Water vapour--infrared difference}, $F_{\mathrm{WV}} = T_{6.2} - T_{11.0}$. The 6.2\,$\mu$m channel is sensitive to moisture in the upper atmosphere, so this difference highlights high clouds and moist upper-level air, which often accompany frontal cloud bands.
    \item \textbf{Split-window difference}, $F_{\mathrm{SW}} = T_{12.0} - T_{11.0}$, which depends on the optical thickness and phase of clouds.
    \item \textbf{Local deviation}, $F_{\mathrm{dev}}(p) = T_{11.0}(p) - \mu_5(T_{11.0})(p)$, which highlights small-scale anomalies within larger cloud systems.
    \item \textbf{Local standard deviation}, $F_{\sigma}(p) = \bigl(\mu_5(T_{11.0}^{2})(p) - \mu_5(T_{11.0})(p)^{2}\bigr)^{1/2}$, which measures cloud texture.
    \item \textbf{Gradient magnitude}, $F_{\nabla} = \bigl((S_x * T_{11.0})^{2} + (S_y * T_{11.0})^{2}\bigr)^{1/2}$, where $S_x$ and $S_y$ are the horizontal and vertical Sobel filters and $*$ denotes convolution. It highlights sharp cloud edges.
    \item \textbf{Hessian determinant}, $F_{\mathrm{H}} = T_{xx}T_{yy} - T_{xy}^{2}$, where $T_{xx}$, $T_{yy}$ and $T_{xy}$ are the second derivatives of $T_{11.0}$ after Gaussian smoothing with a standard deviation of 2 pixels. It highlights curved cloud structures, such as those near occluded fronts.
    \item \textbf{Large-scale smoothing}, $F_{\mathrm{L}} = G_{10} * T_{11.0}$, where $G_{10}$ is a Gaussian filter with a standard deviation of 10 pixels. It retains the outline of large cloud bands and removes small-scale texture.
\end{itemize}

Table~\ref{tab:inputs} lists all input channels.

\section{Additional prediction examples}
\label{app:examples}

Figure~\ref{fig:examples_appendix} shows two additional test samples for both input configurations. The summer sample (22 July 2024) contains several frontal systems over the North Atlantic, and the autumn sample (22 November 2024) contains many occluded fronts. We evaluate predictions only within the UK Met Office region (dashed line), so predictions further east, for example over Russia in the summer sample, are not scored.

\section{Scores at all tolerances}
\label{app:tolerances}

Table~\ref{tab:tolerances} reports the test CSI of SATFRONT for each front class at every tolerance.

\begin{table}[t]
\centering
\caption{Test CSI of SATFRONT for each front class at every tolerance, with UK Met Office labels as ground truth.}
\label{tab:tolerances}
\footnotesize
\setlength{\tabcolsep}{4pt}
\begin{tabular}{@{}llccccc@{}}
\toprule
 & & \multicolumn{5}{c}{\textbf{Tolerance (km)}} \\
\cmidrule(l){3-7}
\textbf{Input} & \textbf{Class} & \textbf{50} & \textbf{100} & \textbf{150} & \textbf{200} & \textbf{250} \\
\midrule
Satellite & Overall & .253 & .348 & .428 & .494 & .548 \\
 & Cold & .270 & .377 & .462 & .529 & .581 \\
 & Warm & .220 & .302 & .378 & .444 & .501 \\
 & Occluded & .267 & .366 & .445 & .508 & .561 \\
\midrule
Satellite+ERA5 & Overall & .374 & .505 & .591 & .649 & .692 \\
 & Cold & .391 & .528 & .616 & .674 & .715 \\
 & Warm & .350 & .479 & .568 & .630 & .674 \\
 & Occluded & .381 & .508 & .588 & .644 & .687 \\
\bottomrule
\end{tabular}
\end{table}

\section{Extraction of frontal analyses from chart images}
\label{app:extraction}

\begin{figure}[h]
\centering
\includegraphics[width=0.8\columnwidth]{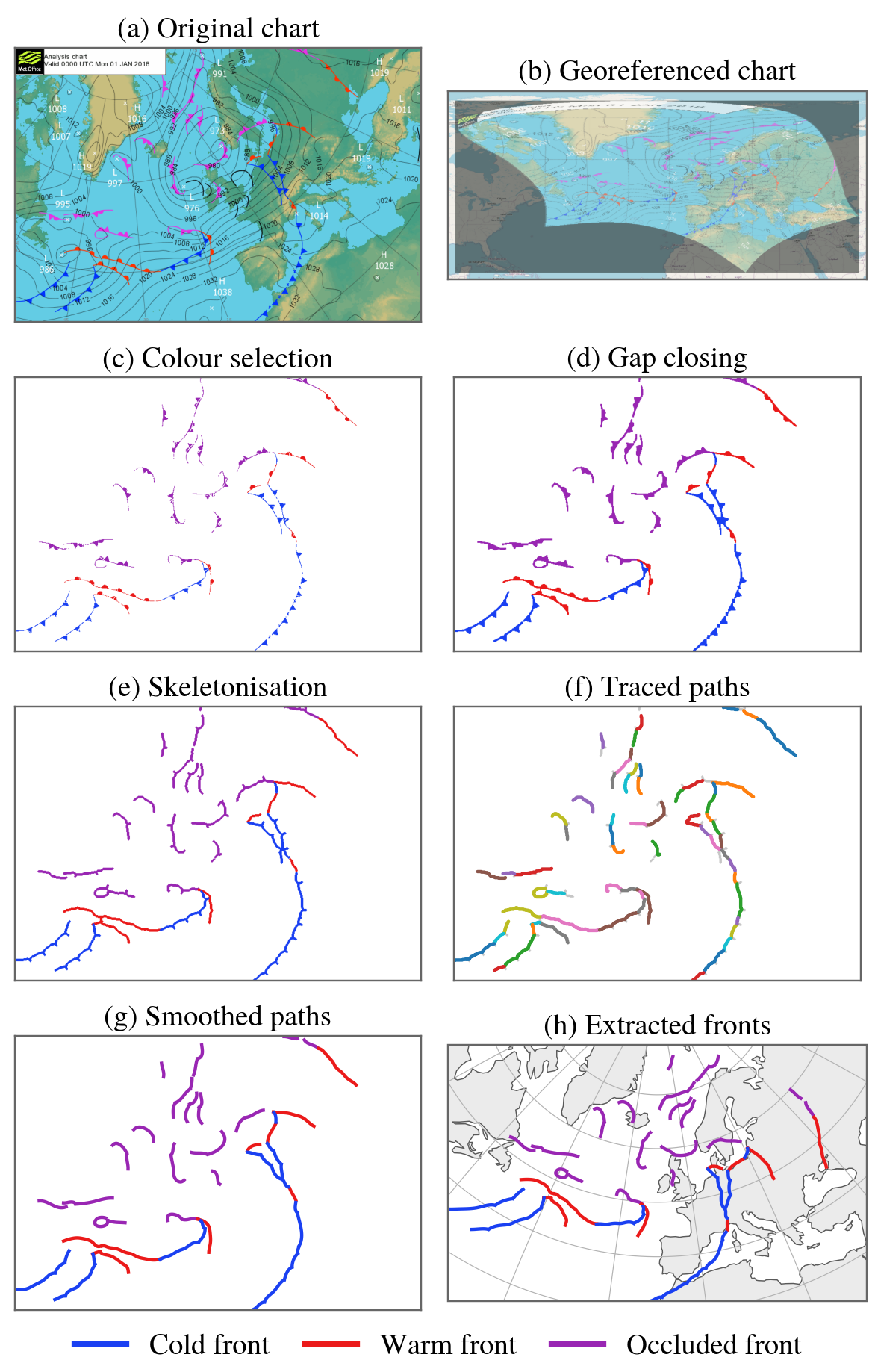}
\caption{Extraction of the fronts from a UK Met Office chart for 1 January 2018, 00:00\,UTC. (a) Original chart. (b) The chart after georeferencing to latitude and longitude with a thin plate spline estimated from manually selected control points. (c) Pixels selected by colour for each front class. (d) The selected pixels after removing isolated pixels and closing small gaps. (e) Skeletons with a width of one pixel, widened for display. (f) Paths traced along the skeletons, each in a different colour. Grey pixels are the short branches formed by the frontal symbols, which are removed. (g) Paths after spline smoothing. (h) The extracted fronts in geographic coordinates. Original chart \textcopyright{} Crown Copyright 2018. Information provided by the National Meteorological Library and Archive -- Met Office, UK. Panels (b)--(h) show our processing of this chart to extract the frontal lines.}
\label{fig:extraction}
\end{figure}

DWD, KNMI, Met \'Eireann, and the UK Met Office publish their analyses only as chart images, each with its own base map, projection, colours, and symbols. We developed a pipeline that converts these images into fronts in geographic coordinates (Figure~\ref{fig:extraction}).

Before processing the charts of a service, we calibrate the mapping from image coordinates to latitude and longitude. We load one chart of the service in QGIS and mark several dozen recognisable points, mostly intersections of latitude and longitude lines, for which we record both the image position and the geographic coordinates. From these control points, we fit either a thin plate spline or a polynomial transformation and keep the one that fits the control points more accurately. Figure~\ref{fig:extraction}b shows the chart in Figure~\ref{fig:extraction}a after georeferencing with a thin plate spline.

\begin{figure*}[b]
\centering
\includegraphics[width=0.65\linewidth]{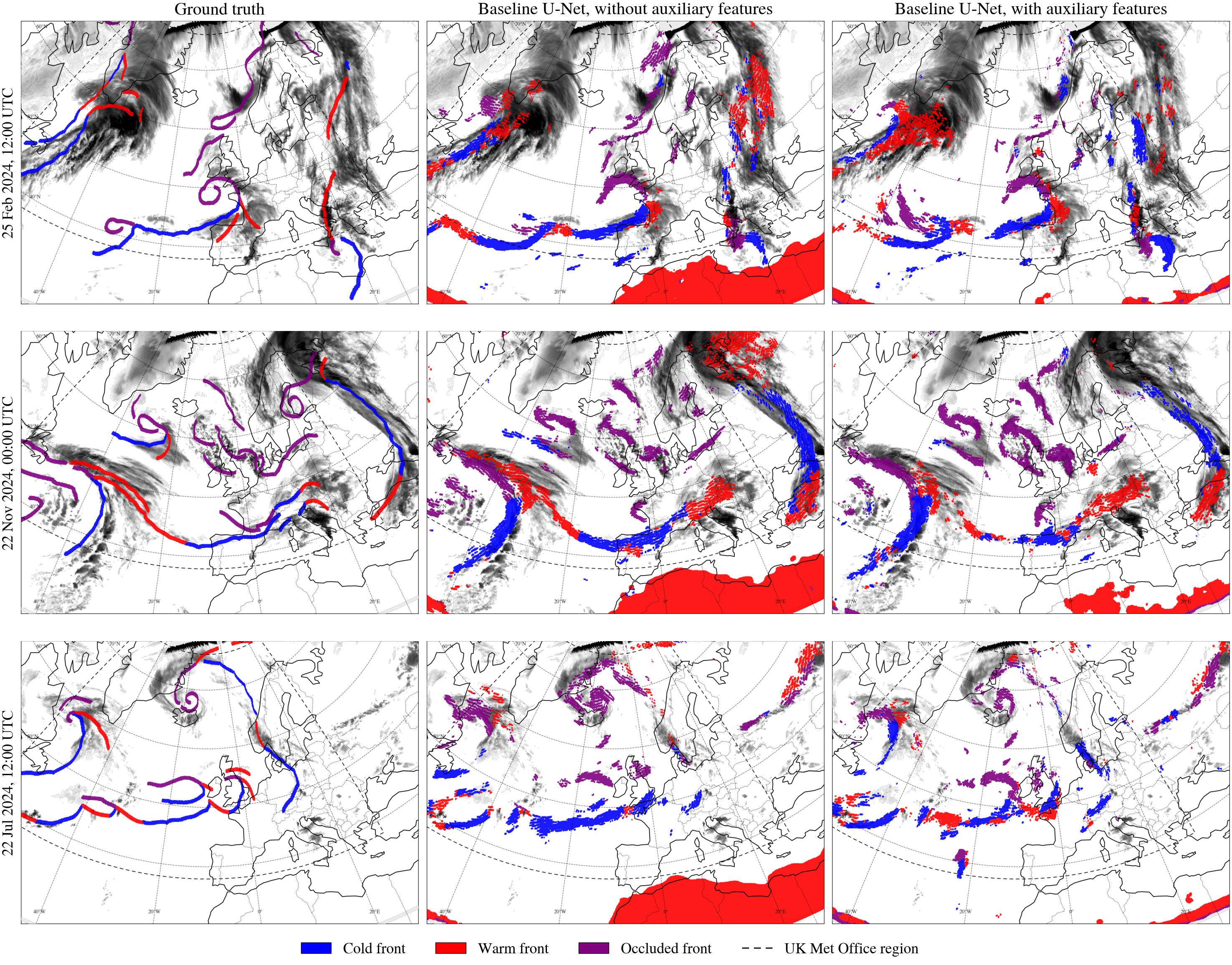}
\caption{Ground truth (left) and satellite-only predictions of the Baseline U-Net without (middle) and with (right) the auxiliary features for three test samples, shown without post-processing. Without the features, the model predicts large spurious warm-front regions over North Africa. With the features, these regions become much smaller but do not disappear. The dashed line marks the UK Met Office region, outside which predictions are not scored and labels do not contribute to the training loss.}
\label{fig:aux}
\end{figure*}

\begin{figure*}[b]
\centering
\includegraphics[width=0.7\linewidth]{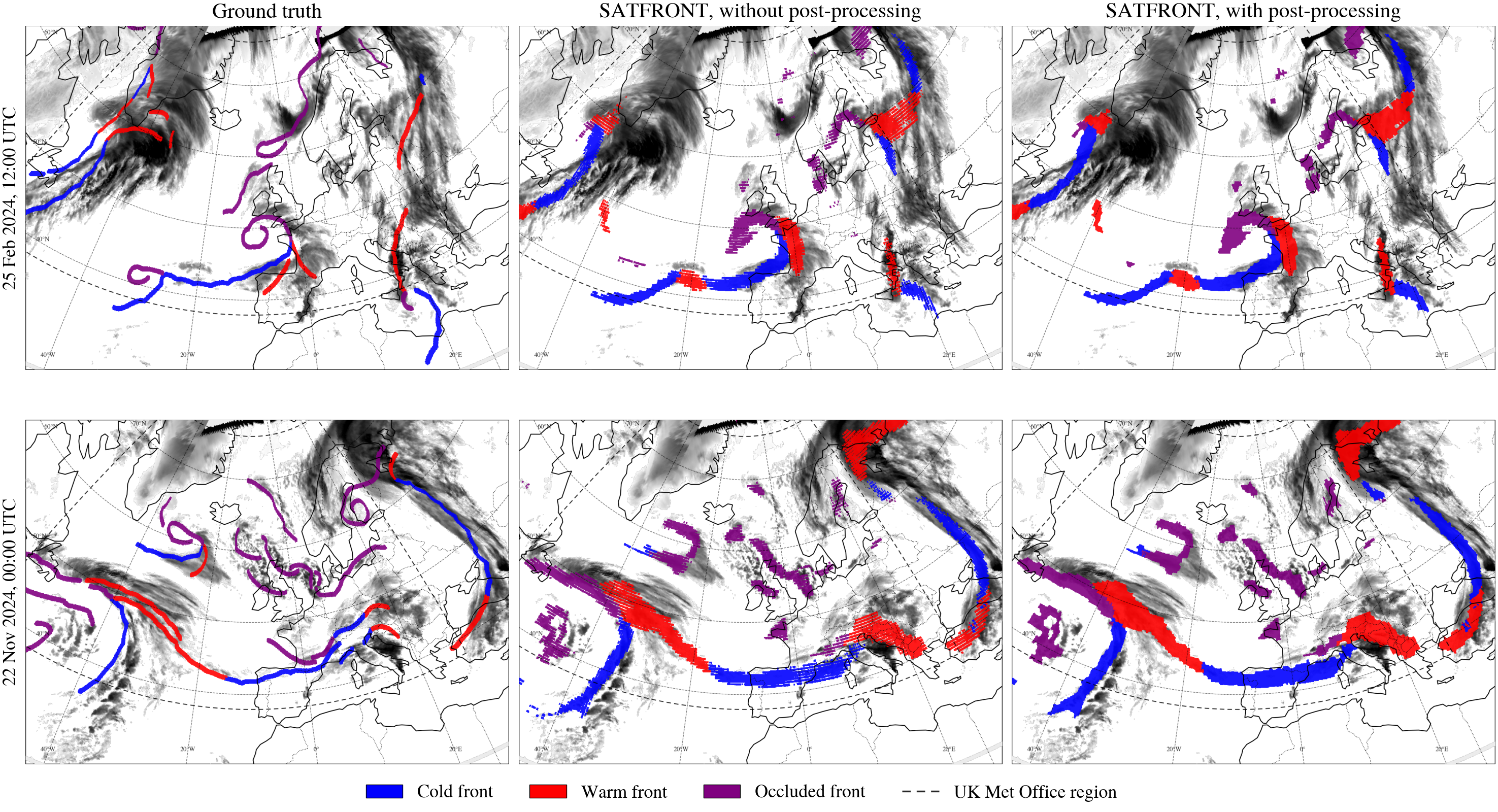}
\caption{Ground truth (left) and satellite-only predictions of SATFRONT without (middle) and with (right) post-processing for two test samples. The dashed line marks the UK Met Office region.}
\label{fig:postprocessing}
\end{figure*}

Each chart (Figure~\ref{fig:extraction}a) then passes through the following steps.
\begin{enumerate}
    \item We select the pixels whose colour lies within a service-specific distance of the colour of each front class, which also assigns each front its class (Figure~\ref{fig:extraction}c).
    \item We remove text, symbols, and other features drawn in the same colours, using connected-component analysis and shape criteria that depend on the service. For the UK Met Office charts, this step removes only isolated pixels.
    \item A morphological closing bridges small gaps in the lines (Figure~\ref{fig:extraction}d), and skeletonisation reduces each line to a width of one pixel (Figure~\ref{fig:extraction}e).
    \item We represent each skeleton as a graph, remove the short branches formed by the frontal symbols, and extract each continuous path as an ordered list of points (Figure~\ref{fig:extraction}f).
    \item We smooth each path with a cubic spline (Figure~\ref{fig:extraction}g), convert it to latitude and longitude with the calibrated transformation, and keep only points at least 0.25$^\circ$ apart (Figure~\ref{fig:extraction}h).
\end{enumerate}
The result is a set of lines with a class and a valid time for each chart.

\section{Selection of the auxiliary features}
\label{sec:aux_selection}

We tested whether the auxiliary features improve the Baseline U-Net trained on UK Met Office labels. With satellite-only inputs, the features did not improve the validation CSI (0.484 without the features and 0.478 with them). With satellite+ERA5 inputs, however, they increased the validation CSI from 0.516 to 0.562. Visual inspection also showed that, with satellite-only inputs, including the features reduced spurious frontal predictions over North Africa (Figure~\ref{fig:aux}). Although this region does not contribute to the training loss or validation scores, the model produces predictions for the full input grid over Europe and the Northeast Atlantic and could potentially be applied to other regions covered by GEO-Ring. The reduction in spurious predictions therefore suggests that the features may improve generalisation beyond the labelled region. One possible explanation is that six of the seven features represent differences between channels or neighbouring pixels, making the model less dependent on absolute brightness temperatures, which can vary substantially between regions, particularly over warm, dry desert surfaces. Based on these results, we include the auxiliary features in both input configurations.

\section{Effect of post-processing}
\label{app:postprocessing}

The raw satellite-only predictions of SATFRONT often break fronts into short, scattered segments with small gaps (Figure~\ref{fig:postprocessing}, middle). These fragments make the frontal structure hard to read, so for visualisation we apply a post-processing step that bridges gaps of up to approximately 12 pixels using a morphological closing and then removes connected regions smaller than 120 pixels (Figure~\ref{fig:postprocessing}, right). We apply this step only to the satellite-only predictions, because the satellite+ERA5 predictions already form continuous fronts. The post-processing is for visualisation only, as in an operational display, and does not affect any reported score, which we always compute on the unprocessed predictions.

\end{document}